\documentclass[12pt]{article}

\def\Journal#1#2#3#4{{#1} {\bf #2}, #3 (#4)}

\def\CQG{\em Class. Quantum Grav.}

\def\PRD{\em Phys. Rev. D }
\def\GRG{\em Gen. Rel. Grav.}

\def\JMP{\em J. Math. Phys.}

\def\CMP{\em Commun. Math. Phys.}

\def\PRL{\em Phys. Rev. Lett.}

\def\ANYAS{\em Ann. N. Y. Acad. Sci.}

\def\NC{\em Nuovo Cimento}

\newcommand{\bm}[1]{\mbox{\boldmath $#1$}}

\def\espaitemps{({\cal V},g)}
\def\varietat{{\cal V}}

\def\xiv{\vec \xi }

\def\lie{{\pounds}}

\def\etaf{\mbox{\boldmath $ \eta $}}

\def\O{\Omega}

\def\f{\varphi}

\def\S{\Sigma}

\def\di{{\rm div}}

\def\be{\begin{equation}}
\def\ee{\end{equation}}
\def\bea{\begin{eqnarray}}
\def\eea{\end{eqnarray}}
\def\bean{\begin{eqnarray*}}
\def\eean{\end{eqnarray*}}

\begin{document}
\title{Novel results on trapped surfaces}
\author{Jos\'e M.M. Senovilla \\
F\'{\i}sica Te\'orica, Universidad del Pa\'{\i}s Vasco, \\
Apartado 644, 48080 Bilbao, Spain \\ 
wtpmasej@lg.ehu.es}
\maketitle 
%\date{\today} 
\begin{abstract} 
A unifying definition of {\em trapped submanifold} for arbitrary codimension 
by means of its mean curvature vector is presented. Then, 
the interplay  between (generalized) symmetries and trapped 
submanifolds is studied, proving in particular that (i) stationary spacetimes 
cannot contain closed trapped nor marginally trapped submanifolds $S$ of 
any codimension; (ii) $S$ can be within the subset where there is a null 
Killing vector only if $S$ is marginally trapped with mean curvature vector 
parallel to the null Killing;  (iii) any submanifold orthogonal to a timelike or null Killing 
vector $\xiv$ has a mean curvature vector orthogonal to $\xiv$.
All results are purely geometric, hold in arbitrary dimension, and
can be appropriately generalized to many 
non-Killing vector fields, such as conformal Killing vectors and the like.
A simple criterion to ascertain the trapping or not of a 
family of codimension-2 submanifolds is given.
A path allowing to generalize the singularity theorems is conjectured
as feasible and discussed. 
\end{abstract} 


\vspace{1cm}
As is well-known, the concept of (closed) trapped surface, first 
introduced by Penrose \cite{P2}, is of paramount importance in many 
physical problems and mathematical developments. In particular, it 
has been an invaluable tool for the development of the singularity 
theorems (see e.g. \cite{P2,HP,HE,S1}), essential for the
analysis of gravitational collapse \cite{P2,HE,GC}, and very useful 
in the study of the cosmic censorship hypothesis \cite{CC} 
or in the numerical evolution of apparently innocuous 
initial data (e.g. \cite{Num} and references therein.) 
This contribution is based partly on \cite{MS} and partly on \cite{S2}, 
where more complete lists of references can be found.

Let $\espaitemps$ be a $D$-dimensional causally orientable 
spacetime with metric $g$ (signature -,+,\dots,+), $\S$ any smooth 
orientable $d$-dimensional manifold, and $\Phi :\S \longrightarrow 
\varietat$ a $C^2$ imbedding. $\S$ and its image 
$\Phi(\S)\equiv S$ in $\varietat$ will be identified 
if no confusion arises. The term `closed' 
is used for {\em compact without boundary} submanifolds. The 
push-forward $\Phi'$ sends a basis of $T(\S)$ to a set of $d$ 
linearly independent vector fields $\vec{e}_{A}$ on $S$ 
($A,B,\ldots =D-d,\ldots, D-1$) which are 
tangent to $S$. Let $\gamma_{AB}=g|_{S}(\vec{e}_{A},\vec{e}_{B})$  
denote the first fundamental form  of $S$ in $\espaitemps$, which is
assumed to be positive definite so that $S$ is 
spacelike. The standard splitting into tangential and normal 
directions to $S$ leads to the typical formula relating the covariant 
derivatives on $\espaitemps$ and $(S,\gamma)$ \cite{Kr,O}:
$$
\nabla_{\vec{e}_{A}}\vec{e}_{B}=\overline{\Gamma}^{C}_{AB}\vec{e}_{C}-\vec{K}_{AB}
$$
where $\overline{\Gamma}^{C}_{AB}$ are the coefficients of the Levi-Civita 
connection $\overline\nabla$ of 
$\gamma$ (i.e. $\overline\nabla\gamma =0$),
and $\vec{K}_{AB}$ is the extrinsic curvature vector of $S$, which is 
orthogonal to $S$. From here it is immediate to obtain the standard 
relation
\be
\forall \bm{v}\in T^{*}(\varietat ), \hspace{1cm}
\Phi^{*}(\nabla \bm{v})=\overline\nabla (\Phi^{*}\bm{v})+\bm{v}(\vec K)
\label{nablas}
\ee
with $\Phi^{*}$ the pull-back of $\Phi$. Equivalently, (\ref{nablas}) can 
be written as ($\mu,\nu,\dots =0,1,\dots ,D-1$)
\be
e^{\mu}_{A}e^{\nu}_{B}\nabla_{\mu}v_{\nu}|_{S}=\overline\nabla_{A} 
\overline{v}_{B}+v_{\mu}|_{S} K^{\mu}_{AB}, \hspace{1cm} 
\overline{v}_{B}\equiv v_{\mu}|_{S}\,\, e^{\mu}_{B}\, .
\label{nablas2}
\ee

For the purposes of this work, the best characterization of trapped surfaces 
is provided by the causal character of their mean curvature vector $\vec H$ 
\cite{S2,Kr}, which is defined as the trace of the extrinsic curvature vector
$$
\vec H \equiv \gamma^{AB}\vec{K}_{AB}\, .
$$
Its virtue is that it can be clearly generalized
to imbedded submanifolds $S$ of any co-dimension, as 
follows\footnote{The zero vector 
$\vec 0$ is null; ``causal vector'' is used for non-spacelike ones.}:
$S$ is said to be
\begin{enumerate}
\item {\em future trapped} if $\vec H$ is timelike and 
future-pointing all over $S$. Similarly for {\em past trapped}.
\item {\em nearly future trapped} if $\vec H$ is causal and 
future-pointing all over $S$, and timelike at least at a point of $S$, 
and correspondingly for {\em nearly past trapped}.
\item {\em marginally future trapped} if $\vec H$ is null and future-pointing 
all over $S$, and non-zero at least at a point of $S$, and analogously for 
{\em marginally past trapped}.
\item {\em extremal} or {\em symmetric} if $\vec H =\vec 0$ all over $S$.
\item {\em absolutely non-trapped} if $\vec H$ is spacelike all over $S$.
\end{enumerate}
In codimension two, points (i), (iv) and (v) coincide with the standard 
nomenclature; point (ii) is a new term coined in \cite{MS}, while 
(iii) is more general than the standard concept (e.g. 
\cite{HE,S1}) ---still, all standard marginally trapped $(D-2)$-surfaces are 
included in (iii)---. On the other hand, the above terminology is 
unusual for the cases $d=D-1$ or $d=1$. 
In the first possibility $S$ is a spacelike hypersurface and thus (i) 
corresponds to the case where its expansion has a sign everywhere on $S$, 
(ii) to the case where it is non-negative or non-positive, (iv) to a maximal $S$, 
while (iii) and (v) are impossible. Similarly, $S$ represents a spacelike curve 
when $d=1$ and therefore the concept of trapping means here that the 
proper acceleration vector of the curve is timelike along $S$ for (i), 
null for (iii), and that $S$ is a geodesic for (iv).

We are now ready to obtain the main formula. To that end, let $\xiv$ be an
arbitrary $C^1$ vector field on $\varietat$ defined on a neighbourhood 
of $S$. Recalling the identity $(\lie_{\xiv} g)_{\mu\nu}=
\nabla_{\mu}\xi_{\nu}+\nabla_{\nu}\xi_{\mu}$ where 
$\lie_{\xiv}$ denotes the Lie derivative with respect to $\xiv$, it trivially 
follows on using (\ref{nablas2}) that
$$
\lie_{\xiv} g|_{S}(\vec{e}_{A},\vec{e}_{B})=\overline\nabla_{A}\overline\xi_{B}+
\overline\nabla_{B}\overline\xi_{A}+2 \xi_{\mu}|_{S} K^{\mu}_{AB}
$$
and taking the trace here 
\be
\fbox{$\displaystyle{\frac{1}{2}
\gamma^{AB}\lie_{\xiv} g|_S (\vec{e}_{A},\vec{e}_{B})=
\overline\nabla_{C}\overline\xi^{C}+ (\xiv\cdot \vec{H})|_{S}}$}
\label{main}
\ee
where $(\, \cdot \,)$ is the $g$-scalar product.

When $\xiv$ is a Killing vector the left hand side of 
(\ref{main}) vanishes. Integrating the right hand side on $S$ and 
using Gauss' theorem the divergence term does not contribute {\em if} 
$S$ is closed, hence ($\etaf_{S}$ is the canonical 
volume element $d$-form associated to $\gamma$)
\begin{center}
$\forall$ closed $S$, and $\forall$ Killing vector $\xiv$: \hspace{1cm}
$\displaystyle{\int_{S} (\xiv\cdot\vec{H}) \, \etaf_{S} =0}$.
\end{center}
Suppose that $\xiv$ is timelike on $S$. If $S$ were (nearly, marginally)
trapped, $(\xiv\cdot\vec{H})$ would have a sign where it is not 
zero on $S$, which is incompatible with the previous result. Thus

{\em There are no marginally trapped, nearly trapped, nor trapped closed 
imbedded submanifolds in any stationary region of $\espaitemps$}. 

As this does not depend on the codimension, in particular one also 
deduces that closed spacelike curves with a future-pointing (or past) acceleration 
vector are forbidden in stationary regions, except for the extremal case of 
closed spacelike geodesics. Similarly, any closed spacelike hypersurface in a 
stationary spacetime must have an expansion which changes sign, 
unless the hypersurface is maximal. As a matter of fact, the
extreme possibility with $\vec H=\vec 0$ is certainly possible for 
all $d$. For instance, asymptotically flat maximal slices always exist in
asymptotically flat stationary spacetimes \cite{C1}, and there are simple 
examples of extremal $S$ of codimension two in stationary spacetimes 
\cite{MS,Kr}.

The previous result might seem intuitively obvious at first, 
it is not straightforward, though, for {\em there are} trapped surfaces in 
stationary, and in static, spacetimes. A simple example taken from 
\cite{S1} (p. 776) is provided by the $(D-2)$-surfaces 
$e^{x^0}-\cosh(x^1)=c_{1},\,\, x^2=c_{2}$ with $c_{1},c_{2}=$const.\ in 
flat Minkowski spacetime with Cartesian coordinates $\{x^{\mu}\}$. Thus, the 
requirement that the submanifolds be closed is indispensable to get 
the result.

Suppose now that the Killing vector $\xiv$ is null and non-zero 
on the closed $S$. Then again if $S$ were trapped, or nearly trapped, 
$(\xiv\cdot\vec{H})$ would be not identically zero and
non-positive (or non-negative) all over $S$ 
in contradiction with the previous result. 
However, $S$ can be marginally trapped if $\xiv$ and $\vec H$ are 
proportional, giving $(\xiv\cdot\vec{H})=0$. Then, 

{\em If $\xiv$ is any Killing vector which is null and non-zero on a closed $S$, 
then $S$ cannot be trapped nor nearly trapped, 
and can be marginally trapped
if and only if its mean curvature vector is parallel to $\vec{\xi}$ everywhere.}

A direct consequence of this statement is that  
pp-wave spacetimes (see e.g. \cite{Exact})
do not admit everywhere expanding (contracting) closed
spacelike hypersurfaces, nor closed trapped or nearly trapped submanifolds.
And any closed marginally trapped submanifold must be contained 
in one of the null hypersurfaces orthogonal to the null covariantly 
constant Killing vector \cite{MS}.

Note that the proofs of the previous results do not need the 
Killing property for $\xiv$: it is enough that just the part of 
$\lie_{\xiv} \, g$ tangent to $S$ is traceless so that the left hand 
side of (\ref{main}) vanishes. This allows for many generalizations, 
among them those using the generalized symmetries introduced recently 
in \cite{CHS}, see the explanations in \cite{MS}. Furthermore, even when 
the part of $\lie_{\xiv} \, g$ tangent to $S$ is not trace-free one can extract 
useful information from formula (\ref{main}). Consider the important example 
of conformal Killing vectors $\lie_{\xiv} \, g=2\phi g$ so that 
$\frac{1}{2}\gamma^{AB}\lie_{\xiv} g|_{S} (\vec{e}_{A},\vec{e}_{B})=\phi d$. 
If $S$ is closed, integrating (\ref{main}) one gets
\begin{center}
$\forall$ closed $S$, and $\forall$ conformal Killing vector 
$\xiv$:\hspace{1cm}
$\displaystyle{\int_{S}\phi \, \etaf_{S} =
\frac{1}{d} \int_{S} (\xiv\cdot\vec{H}) \, \etaf_{S}}$.
\end{center}
It follows that if $\xiv$ is timelike and 
$\phi|_{S}=(1/D)\nabla_{\mu}\xi^{\mu}|_{S}$ does not vanish, the mean 
curvature vector of any (nearly, marginally) trapped closed $S$ 
must be causally oriented in such a way that 
sign$(\xiv\cdot\vec{H})|_{S}$ and sign\,$\phi|_{S}$ coincide where 
they are non-zero ---all timelike homothetic vectors
are here included as they have a constant $\phi$---. Furthermore, extremal 
closed submanifolds are forbidden. As an example one can use the 
timelike conformal Killing vector $\xiv =a(t)\partial_{t}$ of any
Robertson-Walker spacetime, where $a(t)$ is the scale factor and $t$ 
the preferred time coordinate \cite{Exact}. Given that $\lie_{\xiv} g=2\dot{a} g$, 
all possible closed (nearly, marginally) trapped submanifolds must be 
{\em past-trapped} if the Robertson-Walker spacetime is {\em expanding}, 
and vice versa. Besides, in neither case there can be extremal closed 
surfaces, nor maximal closed hypersurfaces; and there cannot be closed 
spacelike geodesics.

Coming back to the general situation, assume now that $S$ is a 
submanifold (closed or otherwise) orthogonal to the vector field $\xiv$. Then 
$\overline\xi_{C}=0$ and formula (\ref{main}) simplifies to 
$\gamma^{AB}\lie_{\xiv} g|_S (\vec{e}_{A},\vec{e}_{B})=2(\xiv\cdot \vec{H})|_{S}$.
Hence, if $\xiv$ is a Killing vector (or any of the afore mentioned 
generalizations), $(\xiv\cdot \vec{H})|_{S}=0$. In particular, 
if $\xiv$ is timelike (respectively null), then any $S$ orthogonal to $\xiv$ 
must be absolutely non-trapped or extremal (resp.\ or marginally 
trapped with $\xiv$ and $\vec H$ parallel), or a mixture thereof. 
These results have special relevance for the case of {\em static} 
regions of the spacetime, where there is a hypersurface-orthoghonal 
timelike Killing vector \cite{MS}.

Combining all the previous results the following statements hold:

{\em In stationary regions of $\espaitemps$, any marginally trapped, 
nearly trapped, or trapped submanifold is necessarily non-closed 
and non-orthogonal to the timelike Killing vector.

In regions with a null Killing vector $\xiv$, all trapped or
nearly trapped submanifolds must be non-closed and 
non-orthogonal to $\xiv$, and any marginally trapped submanifold
must have a mean curvature vector parallel (and orthogonal!)
to the null Killing vector.}

Let us make some comments on why the generalization to arbitrary 
dimension may be important. Apart from the obvious reasons (many 
results can be derived in a simple and direct manner
for closed spacelike curves, or closed spacelike slices, etcetera), 
its relevance arises because it may serve to generalize the 
singularity theorems \cite{P2,HP,HE,S1}
to arbitrary dimension and to many new situations. This application
can be grasped from the following situation. Consider a 4-dimensional 
spacetime with whole cylindrical symmetry \cite{Exact,CSV}, that is to say, 
such that there exist local coordinates in which the line-element reads
\be
ds^2=-A^2dt^2+B^2d\rho^2+C^2d\varphi^2+E^2dz^2, \label{cyl}
\ee
where $\partial_{\f},\partial_{z}$ are spacelike commuting Killing 
vectors whence all the functions depend only on $t$ and $\rho$, and 
the axis is located at $\rho\rightarrow 0$. The coordinate $\f$ is 
closed with standard periodicity $2\pi$. The  
geometrically preferred 2-surfaces are the cylinders given by constant 
values of $t$ and $\rho$, and it is very easy to check whether they 
are trapped surfaces or not: it is enough to compute the causal 
character of the gradient of the function $CE$ ---see below for a 
justification of this. However, these cylinders are {\em not} closed in 
general, so that whether or not they are trapped has no {\em direct} 
relevance for the development of singularities in the sense of 
geodesic incompleteness. 

Nevertheless, the spacelike curves defined by constant values of 
$t,\rho$ and $z$ are certainly {\em closed}, and one wonders if the 
trapping of these curves can lead to similar conclusions, regarding 
incompleteness of causal geodesics, as the trapping of 2-surfaces do 
in the classical singularity theorems. 
However, the theory of points focal to any given spacelike $d$-submanifold along 
causal geodesics is much more involved if $D-d\neq 1,2$, because the variation 
vectors orthogonal to the geodesic are not, {\em all of them}, 
tangent to the submanifold (as opposite to the cases of codimension 1 for 
timelike geodesics and codimension 2 for null ones). Thus, the typical results 
on variation of arc-length \cite{HE,P5,S1} and maximal geodesics are 
unclear, and the focusing effect requires additional restrictions on 
the curvature (not only the null convergence or weak energy conditions).
In spite of all this, there is a kind of underlying common property 
for submanifolds of all codimensions, as has been shown here with the 
unifying definition for them, so that perhaps a \underline{conjecture} along the 
lines of:
{\small ``If appropriate restrictions on the curvature hold and there exists a 
closed future (past) trapped submanifold $S$, then either $E^+(S)$ 
($E^-(S)$) is compact, or the spacetime is null geodesically incomplete 
to the future (past), or both''} may be true.
If this held, then by assuming also the strong causality condition one could 
prove that either $E^+(S)\cap S$ ($E^-(S)\cap S$) is a trapped set, 
or $\espaitemps$ is geodesically incomplete, (see \cite{HP,HE,S1} for 
the definition of $E^{\pm}(S)$, the concept of trapped set, 
and the other conditions mentioned here). The previous 
conjecture is well known to be true for the case of codimension two (and 
one, trivially) by assuming simple ``energy conditions'' \cite{HE,P5,S1}
on the Ricci part of the curvature tensor. If the conjecture became 
concreted, and could thus be proved, then the Hawking-Penrose 
singularity theorem \cite{HP,HE,S1} could be generalized to the case 
of having a closed trapped submanifold of any codimension in 
arbitrary dimension very easily.

This is no place to discuss the conjecture in 
general, but in order to support its plausibility, let us come back to the 
cylindrically symmetric spacetimes. First of all, it is a matter of 
simple calculation to know when the closed curves mentioned before 
are trapped or not, because minus their proper acceleration vector (that 
is, the mean curvature vector) is easily seen to be 
$\vec 
H=C^{-2}(A^{-2}C_{,t}\, \partial_{t}-B^{-2}C_{,\rho}\, \partial_{\rho})$. 
Therefore, $(\vec H\cdot \vec H)=C^{-4}(dC\cdot dC)$, so that the 
causal character of the gradient of {\em only} the function $C$ is 
what matters here. A straightforward consequence is that 
small circles close to the axis are always absolutely non-trapped, 
because if the axis is regular the functions behave as
$C^2 \sim f(t)\rho^2 +O(4)$, $B^2\sim f(t) +O(2)$ and 
$A^2\sim h(t) +O(2)$ when $\rho\rightarrow 0$. More importantly, we 
are now ready to check if the conjecture is valid in some simple 
examples. To that end, take the following particular cylindrically 
symmetric spacetimes, described in their form (\ref{cyl}) by 
\cite{FS,S1}:
$$
A=B=T^2(t)\cosh (3a\rho),\,\,\, 
C= T^2(t)\frac{\sinh(3a\rho)}{3a}\cosh^{-1/3}(3a\rho),\,\,\,
E=T^{-1}(t)\cosh^{-1/3}(3a\rho)
$$
where $a$ is a constant and the function $T(t)$ can be either 
$\cosh(at)$ or $\sinh(at)$. In both cases, the metric is a solution 
of Einstein's field equations for a comoving perfect-fluid energy-momentum 
tensor with realistic equation of state $p=\varrho /3$, being the 
pressure $p=5a^2T^{-4}(t)\cosh^{-4}(3a\rho)$. These spacetimes are 
globally hyperbolic and the dominant energy condition is obviously 
satisfied. An elementary calculation of $dC$ and its causal character
permits to show that the $\varphi$-circles 
are closed absolutely non-trapped curves in the case $T(t)=\cosh(at)$, while 
there always are closed trapped $\varphi$-circles for some values of $t,\rho$ in
the other case $T(t)=\sinh(at)$. This is in agreement with the 
conjecture for, as is known \cite{FS,S1}, the first case 
defines a geodesically complete spacetime while the second one  
is past geodesically incomplete---it has a big-bang singularity at $t=0$---.

Let us finally concentrate on the historically relevant case of codimension 
two. It is well-known that for the case of spherically symmetric 
spacetimes, i.e. such that the line-element can be expressed as
$$
ds^2=g_{ab}(x^c)dx^adx^b +R^2(x^c)d\O^2_{D-2} \, ,
$$
where $\{x^a\}$ ($a,b,\dots =0,1$) are coordinates on the Lorentzian 
2-dimensional metric $g_{ab}$ and $d\O^2_{D-2}$ is the round metric on 
the $(D-2)$-sphere, the causal character of the gradient of the 
function $R$ fully characterizes the trapping or not of the spheres, 
and strongly affects that of any other closed submanifold of codimension 2
(e.g. \cite{S2}). The question arises of how to generalize this simple criterion 
to arbitrary spacetimes. This was studied in \cite{S2}, where the 
following simple way to check the trapping of surfaces was presented. 
Take any given family of $(D-2)$-dimensional spacelike surfaces $S_{X^a}$ 
described by $\{x^a=X^a\}$ where $X^a$ are constants
and $\{x^{\alpha}\}$ are local coordinates in $\varietat$. 
The line-element can be locally written as
\begin{equation}
ds^2=g_{ab} dx^adx^b+2g_{aA}dx^adx^A+g_{AB}dx^Adx^B
\label{sdg}
\end{equation}
where $\det g_{AB} >0$ and all the $g$-components can depend on 
{\em all} coordinates: $g_{\mu\nu}(x^{\alpha})$. Define
$$
e^U\equiv +\sqrt{\det g_{AB}} , \hspace{2cm} \bm{g}_{a}\equiv 
g_{aA}dx^A \label{U}\, .
$$
Then, the mean curvature one-form of $S_{X^a}$ reads ($f_{,a}= \partial_{x^a} f$)
\be
\fbox{$\displaystyle{
H_{\mu}=\delta^a_{\mu}\left(U_{,a}-\di \vec{g}_{a} \right)\vert_{S_{X^a}}}$}
\label{H}
\ee
where $\di$ is the divergence operating on vectors at each $S_{X^a}$, 
so that its Lorentz length is
\begin{equation}
g^{\mu\nu}\vert_{S_{X^a}}H_{\mu}H_{\nu}=g^{bc}\vert_{S_{X^a}}H_{b}H_{c}
=\left.g^{bc}\left(U_{,b}-\di \vec{g}_{b} \right)
\left(U_{,c}-\di \vec{g}_{c} \right)\right\vert_{S_{X^a}}\, .
\label{kap}
\end{equation}
(Notice that $g^{bc}$ is not necessarily the inverse matrix of $g_{bc}$!).
Contracting (\ref{H}) with a vector field $\xiv$ one 
obtains an expression for $(\xiv\cdot \vec H)$ which can be related, 
after some calculations, to (\ref{main}) (when 
$d=D-2$). Note also that one only needs to compute
the norm of $\bm{H}$
{\it as if it were a one-form in the 2-dimensional metric $g^{ab}$}. 
Thus, (\ref{kap}) provides the sought criterion. Many applications 
arise, for a selection of them the reader is referred to \cite{S2}.

\section*{Acknowledgements} This contribution is largely based on a 
collaboration with Marc Mars, to whom I am also grateful for many 
comments. Financial support under
grants BFM2000-0018 of the Spanish CICyT and 
no. 9/UPV 00172.310-14456/2002 of the University of the Basque 
Country is acknowledged.

\end{document}